\documentclass[11pt]{article}
\usepackage{amsmath,amssymb,fullpage,graphicx}
\usepackage{algorithm,algpseudocode}
\usepackage{caption,subcaption}
\usepackage{hyperref}
\usepackage[square,sort,comma,numbers]{natbib}
\usepackage{indentfirst}
\usepackage{xcolor}

\title{Weakest link pruning of a dendrogram}
\author{Jiacheng Ge
\footnote{Department of Statistics, Stanford Univ.}, and Robert Tibshirani\footnote{Department of Departments of  Biomedical Data Science and of Statistics, Stanford Univ. }  \\
}

\begin{document}
\maketitle

\begin{abstract}
\end{abstract}
Hierarchical clustering is a popular method for identifying  distinct groups in a  dataset. The most commonly used method for pruning a dendrogram  is via a single horizontal cut. In this paper, we propose a new  technique ``weakest link optimal pruning''. We  prove its superiority over horizontal pruning and provide some examples illustrating how the two methods can behave quite differently.\

\section{Introduction}
Hierarchical clustering is a  widely used method for identifying distinct groups in data in an unsupervised setting. There are two main types of hierarchical clustering strategies: agglomerative (bottom-up) and divisive (top-down). Both strategies use a greedy approach and  we focus on agglomerative clustering in this paper. One benefit of hierarchical clustering is that any legitimate distance metric can be applied. Common distance metrics include average linkage, single linkage, and complete linkage. The hierarchy of clusters is usually presented in a dendrogram with desired clusters as its branches. The height of a node is proportional to its within-cluster dispersion. ``Pruning'' the dendrogram involves the removal of some of its lower branches. 

The most popular dendrogram pruning method is the horizontal cut: we merge the leaves into several clusters either by specifying the desired number of clusters or the cut height. Every contiguous branch of objects below the cut is grouped together. This technique is a straightforward, simple method with many appealing qualities. 

Recall that a dendrogram is constructed from the bottom up, by merging the closest pair of nodes at each stage, and it is displayed
with the height of each node equal to the distance at which the merge occurs. From this, one might expect  that the horizontal cut yields the  tightest set of clusters for a given number of leaves. Due to the greedy nature of the merges, however, this turns out not to be the case,
as we show here.

Dynamic Tree Cut (Langfelder et al.) \citep{dynamic} is a dendrogram pruning method based on investigating the dendrogram's branch shapes. It has two variants: “Dynamic Tree” cut and “Dynamic Hybrid” cut. The Dynamic Tree cut is a top-down approach that involves adaptively breaking down and recombining clusters until a stable number of clusters is reached. The Dynamic Hybrid cut, on the other hand, is a bottom-up process that involves identifying branches that meet certain criteria and then assigning unassigned objects to the clusters found in the first step if they are sufficiently close. Dynamic tree cut is flexible and has been found to work well on bioinformatic tasks. However, it is computationally intensive and hard to interpret. Also, it is still unclear how to select the height and shape parameters and how to estimate the number of clusters.

Dynamic programming, which was created by Richard Bellman \cite{dp2}, is a technique that is used in both mathematical optimization and computer programming to solve complex problems by breaking them down into simpler sub-problems and solving them recursively. We aim to minimize a chosen loss when pruning a dendrogram into $k$ groups. While it is possible to enumerate all combinations, a more efficient method is to identify the optimal subtrees by dynamic programming \cite{dp1}. To optimally cut a tree into $k$ clusters, it equivalently determines the optimal number of groups in the left and right branches of the root ($k_l$ and $k_r$, respectively), with the condition that $k_l + k_r = k$.  This process is then repeated for each branch, recursively until it reaches the leaves of the dendrogram or all remaining nodes fall into one cluster. An important drawback of this method is that it does not necessarily produce a nested sequence of subtrees.

In Section \ref{sec:two}, we propose weakest link optimal pruning and prove theoretically its advantage over the horizontal pruning method. In this paper, we mainly focus on the comparison between the horizontal cut and our optimal cut.  We provide some examples in Section \ref{sec:three}, including simulated examples with and without clusters, and some popular real datasets, illustrating how the two methods behave differently. In Section \ref{sec:class}, we  embed the two pruning methods into a classification task using a real dataset and compare their performances.  We combine the optimal pruning
sequence with the Gap test in Section \ref{sec:gap} to estimate the ideal number of clusters. Finally, Section \ref{sec:discussion} contains some discussion.

\section{Our proposed method: weakest link pruning}
\label{sec:two}

\subsection{Set up}
Our approach follows closely the weakest link pruning method used in
classification and regression trees (CART) \citep{CART}, a supervised learning method. In our unsupervised clustering problem,  the only change we make here is in the loss function, where  we replace misclassification loss with the within-cluster dispersion loss.

Let $T$ be a dendrogram and $t$ be a node in $T$. If there is only one observation in a node $t$, we call it a terminal node or leaf. Let $\tilde{T}$ denote all terminal nodes in $T$. Denote by $D(x_i, x_i')$ the distance between
points $x_, x_i'$: most commonly we use squared error loss $||x_i-x_i'||^2$. The loss function $R(t)$ is defined as the sum of the distances between all possible pairs of data points in node $t$, i.e. $$R(t) = \underset{x_i, x_{i'} \in  t}{\sum} D(x_i, x_{i'}).$$

The loss function $R$ for $T$ is defined as the total dispersion loss within all leaves, i.e. $$R(T) = \underset{t \in \tilde{T}}{\sum} R(t) = \underset{t \in \tilde{T}}{\sum} \, \underset{x_i, x_{i'} \in  t}{\sum} D(x_i, x_{i'}).$$

Let $T_{max}$ denote an unpruned dendrogram with only one observation in each terminal node. If $T$ is a subtree of $T_{max}$, we denote $T \subseteq T_{max}$. Our goal is to find the ``optimal'' dendrogram $T \subseteq T_{max}$. However, $R(T)$ alone is not a useful metric for finding the ``optimal'' subtree because it always prefers larger trees. It is easy to see that $R(T)=0$ when each leaf only contains one case. So, we introduce a complexity penalty to control the balance between dispersion and tree size. For any subtree  $T \subseteq T_{max}$, we define its complexity as $|\tilde{T}|$, the number of terminal nodes in $T$. Let $\alpha \geqslant 0$ be a real number and called the complexity parameter. We define the cost-complexity measure $R_{\alpha}(T)$ as:
$$R_{\alpha}(T) = R(T) + \alpha|\tilde{T}|.$$

Our general approach is to (a) for each $\alpha$, find the subtree
that minimizes $R_{\alpha}(T)$  and then (b) estimate the best value
 $\hat\alpha$ and hence the optimal subtree $T(\hat\alpha)$ using a method for estimating the best number of clusters. In Section \ref{sec:gap} below we illustrate this using the Gap Test, but emphasize that other methods for estimating the number of clusters could be used.

For (a), our goal is to find subtree $T(\alpha) \subseteq T_{max}$ which minimizes $R_{\alpha}(T)$; $T(\alpha)$ is called the minimizing subtree for $\alpha$, i.e. $$R_{\alpha}(T(\alpha))= \underset{T \subseteq T_{max}}{\text{min}} R_{\alpha}(T).$$ The smallest such $T(\alpha)$, in terms of complexity, is defined as the smallest minimizing subtree for the complexity parameter $\alpha$. By its definition, the smallest minimizing subtree for a given $\alpha$ is unique.

In this paper, we use squared distance as the dispersion loss, i.e. $$D(x_i, x_{i'}) = \| x_i - x_{i'}\|^2,$$ then $$R(t) = \underset{x_i, x_{i'} \in  t}{\sum} \| x_i - x_{i'}\|^2 \quad {\rm and} \quad R(T) = \underset{t \in \tilde{T}}{\sum} R(t) = \underset{t_j \in \tilde{T}}{\sum} \, \underset{x_i, x_{i'} \in  t_j}{\sum} \| x_i - x_{i'}\|^2.$$

\subsection{Weakest link optimal pruning}
 
To find the smallest minimizing tree, we adapt weakest link pruning: we successively collapse the internal node that causes the smallest per-node rise in $R(T)$ until a single-node (root) dendrogram is produced. Given a dendrogram of $n$ leaves, choosing to cut the internal node with the smallest per-node increase in $R(T)$ is equivalent to finding the subtree with fewer than $n$ leaves which has the smallest value of $R$. By performing  weakest link pruning, we generate a sequence of minimal cost-complexity subtrees. Notably, the subtrees are decreasing in their complexity, but their number of leaves may not be consecutive. In other words, some tree sizes could be skipped.

On pages 66-71 of CART book \citep{CART},  the following facts are established:
\begin{enumerate}
 \item For every value of $\alpha$, there exists a (unique) smallest minimizing subtree. 
 \item The weakest link sequence of minimal cost-complexity trees contains the smallest minimizing subtree.
\end{enumerate}
The proofs of these claims follow directly from the CART results \citep{CART}.

\subsection{Optimal pruning versus horizontal pruning}
First, we claim that weakest link pruning always yields a total dispersion loss equal to or less than the horizontal cut for a given number of terminal nodes. We will prove this by contradiction. Let $T_{max}$ be a dendrogram and $\{T_{s_i}\}$ be its sequence of minimal cost-complexity subtrees. Assume that there exists a horizontally pruned dendrogram $T_h \subseteq T_{max}$ that is not in $\{T_{s_i}\}$ and the dendrogram has a smaller loss than any of the optimal pruned subtrees, i.e. $R(T_1) \leqslant\underset{s_i}{{\rm min}}R(T_{s_i})$. The horizontal cut is denoted by $h_1$. All the nodes equal to or below $h_1$ are already collapsed while all nodes above are not. Denote the pruned dendrogram before $h_1$ happening by $T_{h'}$. Hence:
\begin{itemize}
    \item If all the nodes above the $h_1$ have a larger per-node rise in $R(T_{h'})$ than all the nodes equal to or below $h_1$ when being collapsed, then by the definition of weakest link pruning, $T_1$ must be in $\{T_{s_i}\}$.
    \item If there exists a node $t_1$ above $h_1$ and a node $t_2$ equal to or below $h_1$ such that per-node rise of $R(T_{h'})$ when cutting $t_1$ is smaller than cutting $t_2$, then we can reduce $R(T_h)$ by collapsing $t_1$ and expanding $t_2$. 
\end{itemize}
In both cases, there's a contradiction to our assumption. Therefore, the horizontal cut solution is either in the weakest link merge sequence or the resultant subtree has a higher loss, implying that weakest link pruning is never worse than horizontal cut for a fixed number of leaves.

Second, recall that a sequence of minimal cost-complexity subtrees may skip some tree sizes. Without loss of generality, let's assume the tree complexities in the optimal subtree sequence are ($n, n-1, ..., k+c+1, k+c, ..., k, ..., 1$). When we have a subtree of $k+c+1$ nodes, we have the option to prune it into k+c, $k+c-1, ..., k$ leaves. Since we choose the next merge by comparing the smallest per-node increase in $R$, then the best subtree with $k+c-1, k+c-2, ..., k+1$ leaves must have bigger losses than the best subtree with $k+c$ leaves or $k$ leaves. By transitivity, the horizontal pruned trees with $k+c-1, k+c-2, ..., k+1$ leaves will have even larger losses than the best minimal cost-complexity subtree with $k+c$ leaves or $k$ leaves.

To summarize, given a horizontally pruned dendrogram $T_h \subseteq T_{max}$, we can always find a $T$ in the sequence of minimal cost-complexity subtrees of $T_{max}$ such that $R(T_h) \geqslant R(T)$. This implies that horizontal pruning can never do better than weakest-link optimal pruning.

\section{Some examples}
\label{sec:three}
\subsection{A simple example}
\begin{figure}[!ht]
\centering
\begin{subfigure}{.5\textwidth}
  \centering
  \includegraphics[scale = 0.6]{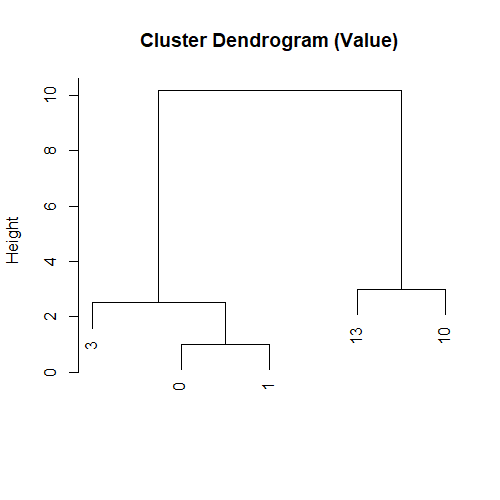}
  \caption{\em Leaves labelled by their data values}
  \end{subfigure}%
\begin{subfigure}{.5\textwidth}
  \centering
  \includegraphics[scale = 0.6]{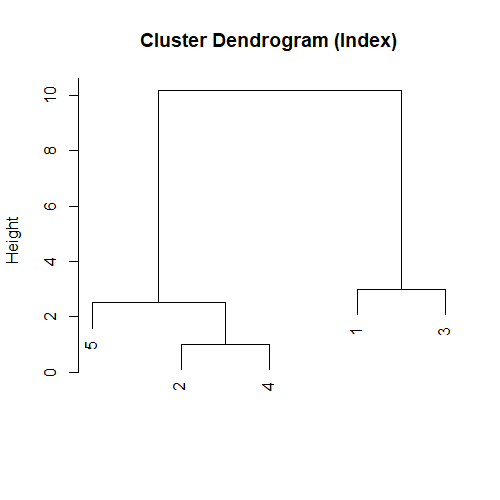}
  \caption{\em Leaves labelled  by the node numbers 1--5}
\end{subfigure}
\caption{\em Cluster dendrograms for a simple example}
\label{fig:dend ex}
\end{figure}

Here we illustrate that the optimal pruning and horizontal cuts can behave differently. Consider a dataset of five (scalar) data points: 13, 0, 10, 1, 3, indexed by 1, 2, 3, 4, 5. We created a dendrogram for the dataset via average linkage, which is shown in Figure \ref{fig:dend ex}. The left plot labels the leaves by their data values and the right one shows the node numbers 1--5.  

\begin{figure}[!ht]
\centering
\begin{subfigure}{.5\textwidth}
  \centering
  \includegraphics[scale = 0.5]{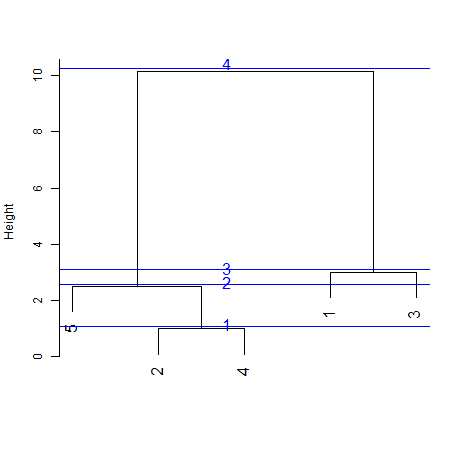}
  \caption{\em Cuts in horizontal pruning}
\end{subfigure}%
\begin{subfigure}{.5\textwidth}
  \centering
  \includegraphics[scale = 0.5]{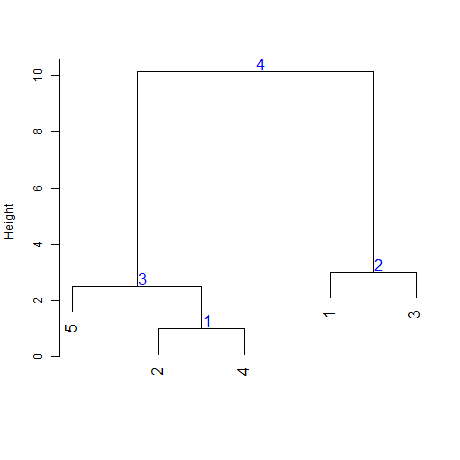}
  \caption{\em Cuts in optimal pruning}
\end{subfigure}
\caption{\em The cuts in two different pruning processes}
\label{fig:label merge}
\end{figure}

The left plot of Figure \ref{fig:label merge} shows the cuts in the horizontal pruning process using the {\tt cutree} function in the R language package {\tt stats} \citep{stats}. This process collapsed the lowest internal node at each step from bottom to top. The right plot labels the merges in the optimal pruning process. Notice that the cuts were not increasing in height at each step. (It is worth mentioning that it's also common for some internal nodes to be skipped.)

\begin{figure}[!ht]
\centering
\begin{subfigure}{.5\textwidth}
  \centering
  \includegraphics[scale = 0.38]{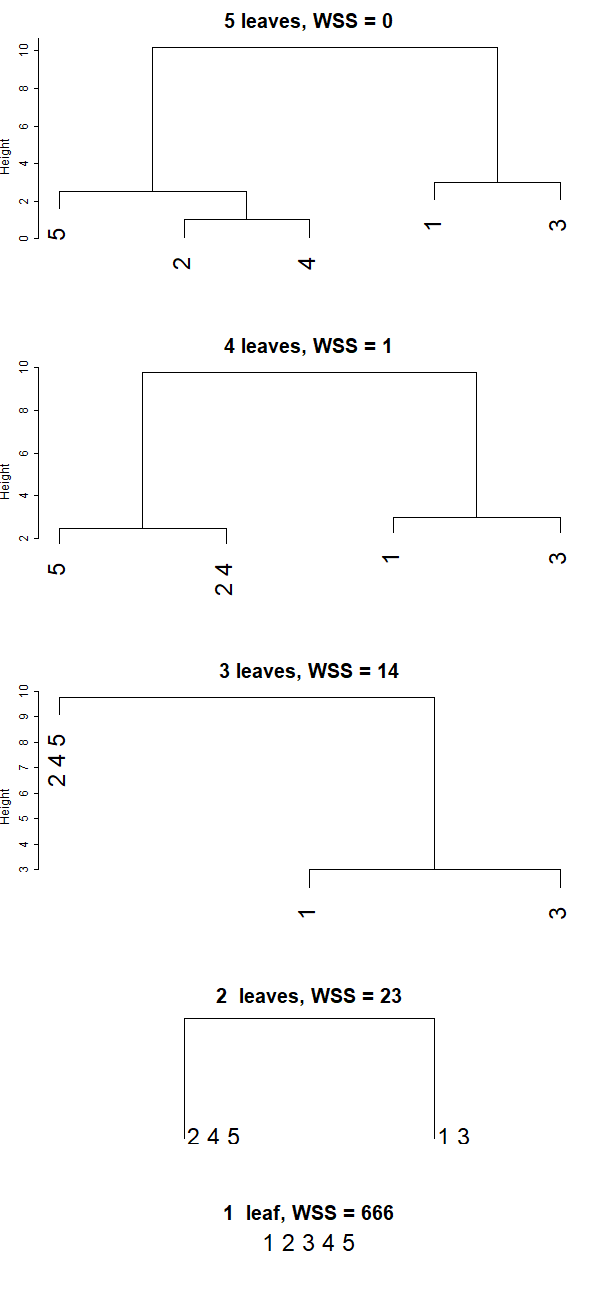}
  \caption{\em Horizontal pruning process}
\end{subfigure}%
\begin{subfigure}{.5\textwidth}
  \centering
  \includegraphics[scale = 0.38]{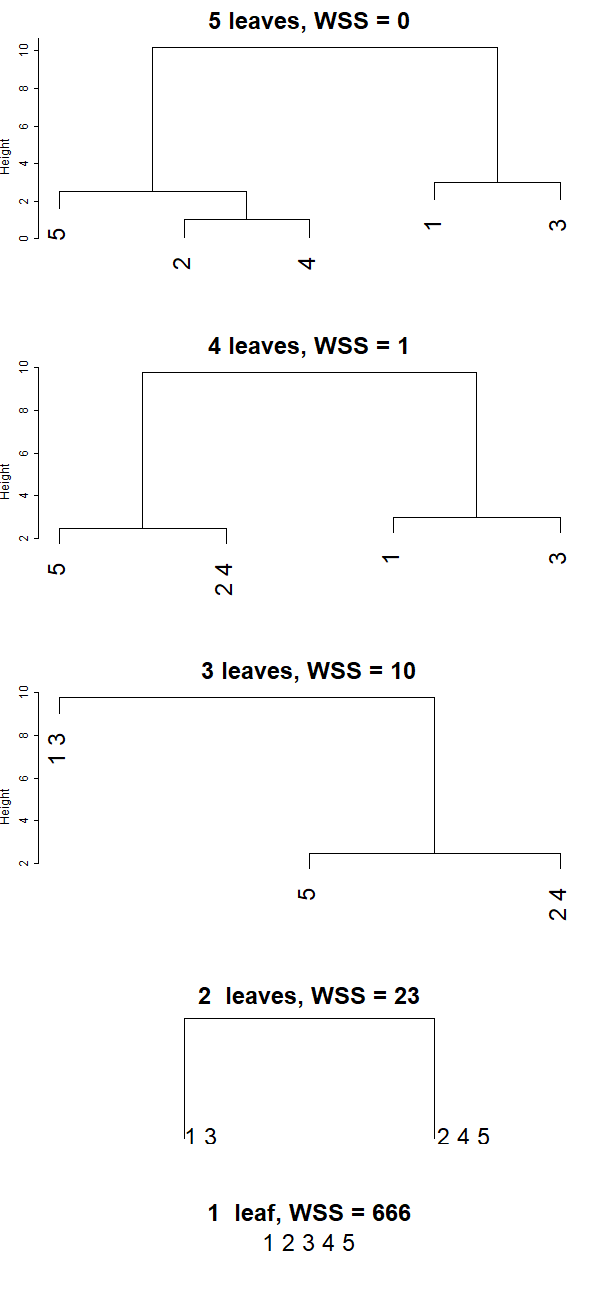}
  \caption{\em Optimal pruning process}
\end{subfigure}
\caption{\em The sequence of subtrees from two different pruning strategies.}
\label{fig:pruned tree}
\end{figure}

Figure \ref{fig:pruned tree} details the subtrees produced by two different pruning methods. Notably, resulting  subtrees with 5, 4, 2, and 1 terminal nodes were the same for both methods. However, optimal pruning merged node 1 and node 3 before merging node (2 4) and node 5. And its resulting within-cluster sums of squares (WSS) is 10 when there were 3 leaves, smaller than the value 14 for the  horizontal cuts. 

\subsection{Experiments with simulated datasets}
\subsubsection{Random datasets with no clustering}
\begin{figure}[!ht]
    \begin{centering}
    \includegraphics[scale=0.6]{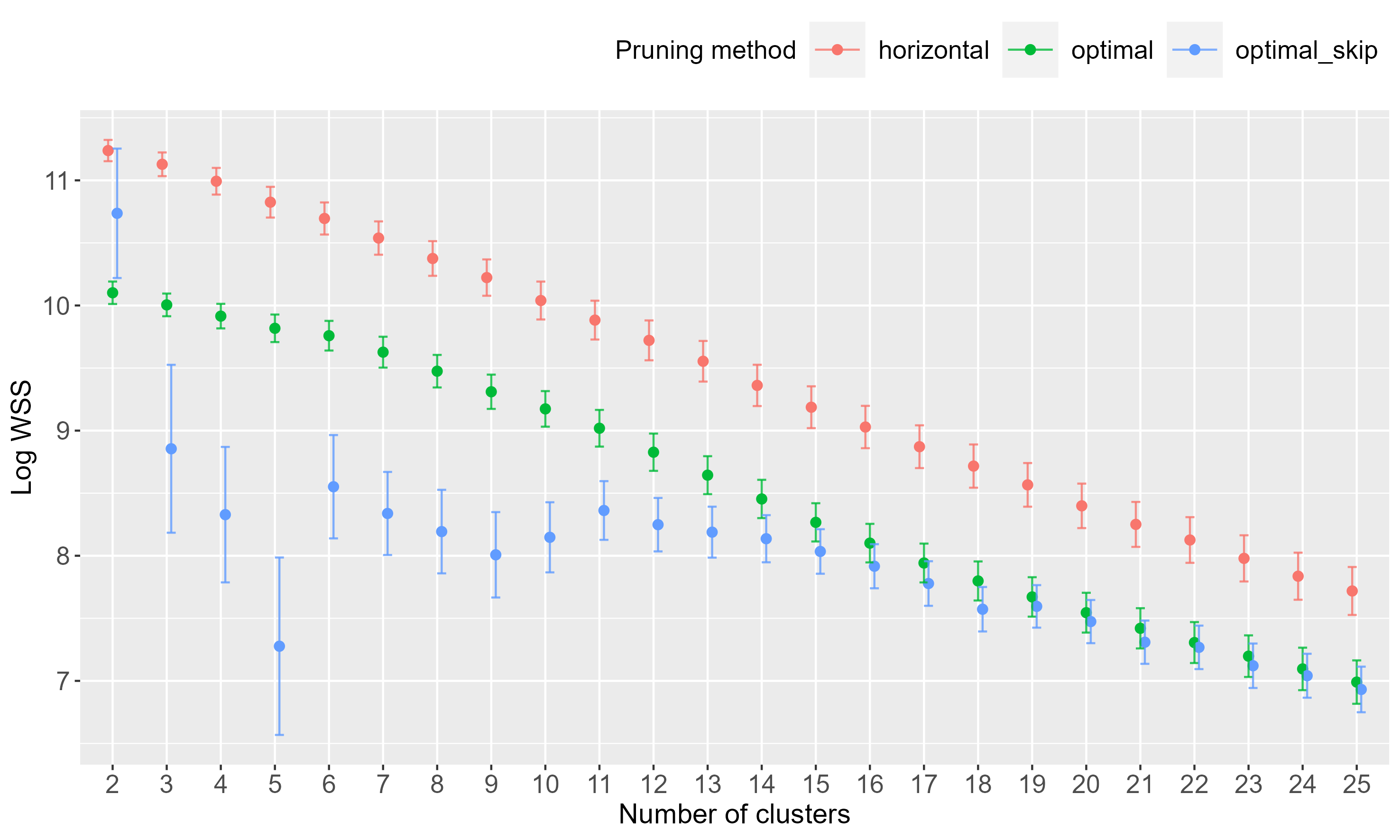}
    \caption{\em Log WSS mean and SE comparison of the two pruning methods on 200 random datasets with no clusters. ``Optimal pruning'' increases the number of leaves to the closest available number if the subtree with a certain number of leaves is not in the sequence of optimal pruned subtrees, while ``Optimal\_skip'' skips the subtree.} 
    \label{vs1}
    \end{centering}
\end{figure}

We randomly generated 200 datasets $\mathit{X_i}$ with number of observations $n_i$  between 30 and 100, and number of features $p_i$ between 1 and 50. We chose $n_i$ and $p_i$ uniformly sampled and $x_{ij}$ $\overset{\mathrm{iid}}{\sim}$ $\mathcal{N}(0, \mathit{I_{p_i}})$. Then, we ran the two pruning methods, horizontal and optimal weakest-link pruning, on the 200 sets of data for comparison. Figure \ref{vs1} shows the log of  WSS when the dendrogram was pruned down to between 2 and 25  terminal nodes. Each colored circle represents the mean of the (log) WSS and the error bar represents one standard error. For optimal pruning, if the subtree with a certain  number of leaves was not in the sequence of optimal pruned subtrees, we increased the number of leaves to the closest available number. For the optimal\_skip pruning, if the subtree with a certain number of leaves was not in the sequence of optimal pruned subtrees, we skipped this subtree, thus the loss is defined as NA. 

We see that the optimal cut has consistently smaller WSS, sometimes by a large margin. For some specific numbers of clusters, the variance of the optimal\_skip WSS is higher because some number of clusters are skipped and the amounts of losses are reduced. 

\begin{figure}[!ht]
    \begin{centering}
    \includegraphics[scale=0.6]{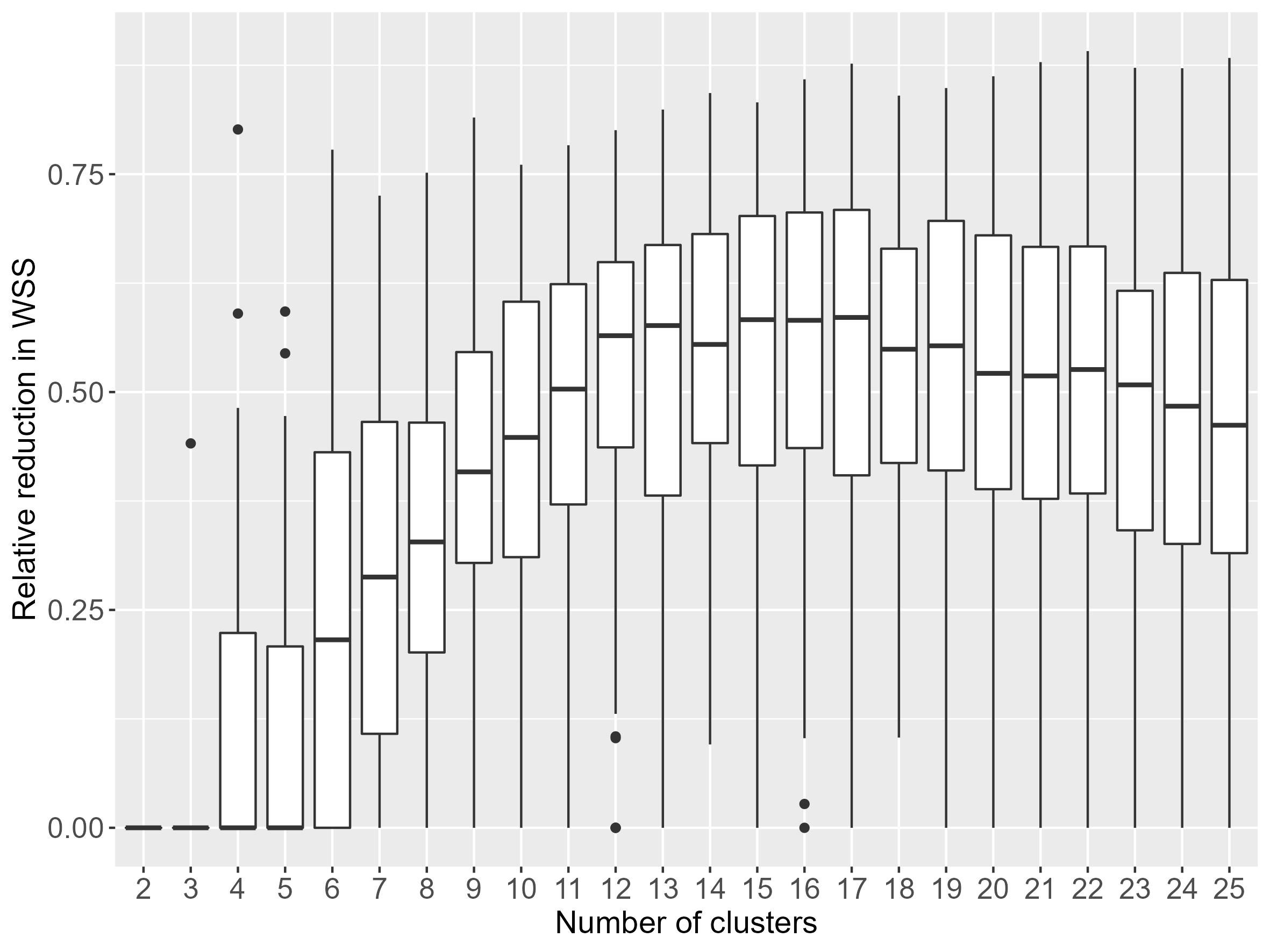}
    \caption{\em Relative reduction in WSS, for optimal pruning versus horizontal cut on 200 random datasets with no clustering.} 
    \label{diff1}
    \end{centering}
\end{figure}

Figure \ref{diff1} shows the relative improvement in the loss for weakest-link pruning relative to horizontal cut. In this plot, the relative differences were only calculated when the subtree with the  clusters was a member of the sequence of minimizing subtrees. The loss reduction is quite remarkable and the median of loss reduction (after removing NAs due to skipping numbers of clusters) was 0.51. As demonstrated by both plots, optimal pruning shows better performance than the horizontal cut when the data is completely random.

\subsubsection{Random datasets with well-separated clusters}
\begin{figure}[!ht]
    \begin{centering}
    \includegraphics[scale=0.4]{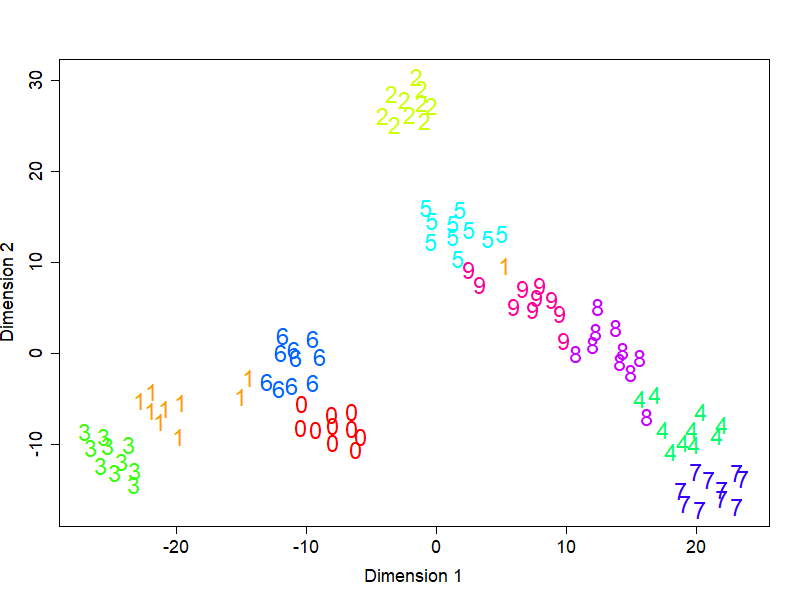}
    \caption{\em TSNE projection of a random dataset with well-separated clusters: 100 points, 15 dimensions, 10 clusters} 
    \label{fig:clus_ex}
    \end{centering}
\end{figure}

We randomly generated 200  datasets  with number of observations $n_i$  between 30 and 100, number of features $p_i$ between 1 and 50, and the numbers of clusters $c_i$ between 3 and 15, all uniformly sampled. We generated $x_{ij}$ $\overset{\mathrm{iid}}{\sim}$ $\mathcal{N}(0, \mathit{I_{p_i}})$ first, and then shifted the data from each cluster by a vector of length $p_i$ whose components repeat an integer uniformly sampled from 1 to $c_i$ without replacement. The sizes of all clusters were equal, except for the last cluster which is adjusted for indivisibility. The 2D projection of an example dataset produced by {\tt Rtsne} R package \citep{Rtsne} is illustrated in figure \ref{fig:clus_ex}. We can see that the clustered data are well separated with a small amount of overlap.

\begin{figure}
    \begin{centering}
    \includegraphics[scale=0.6]{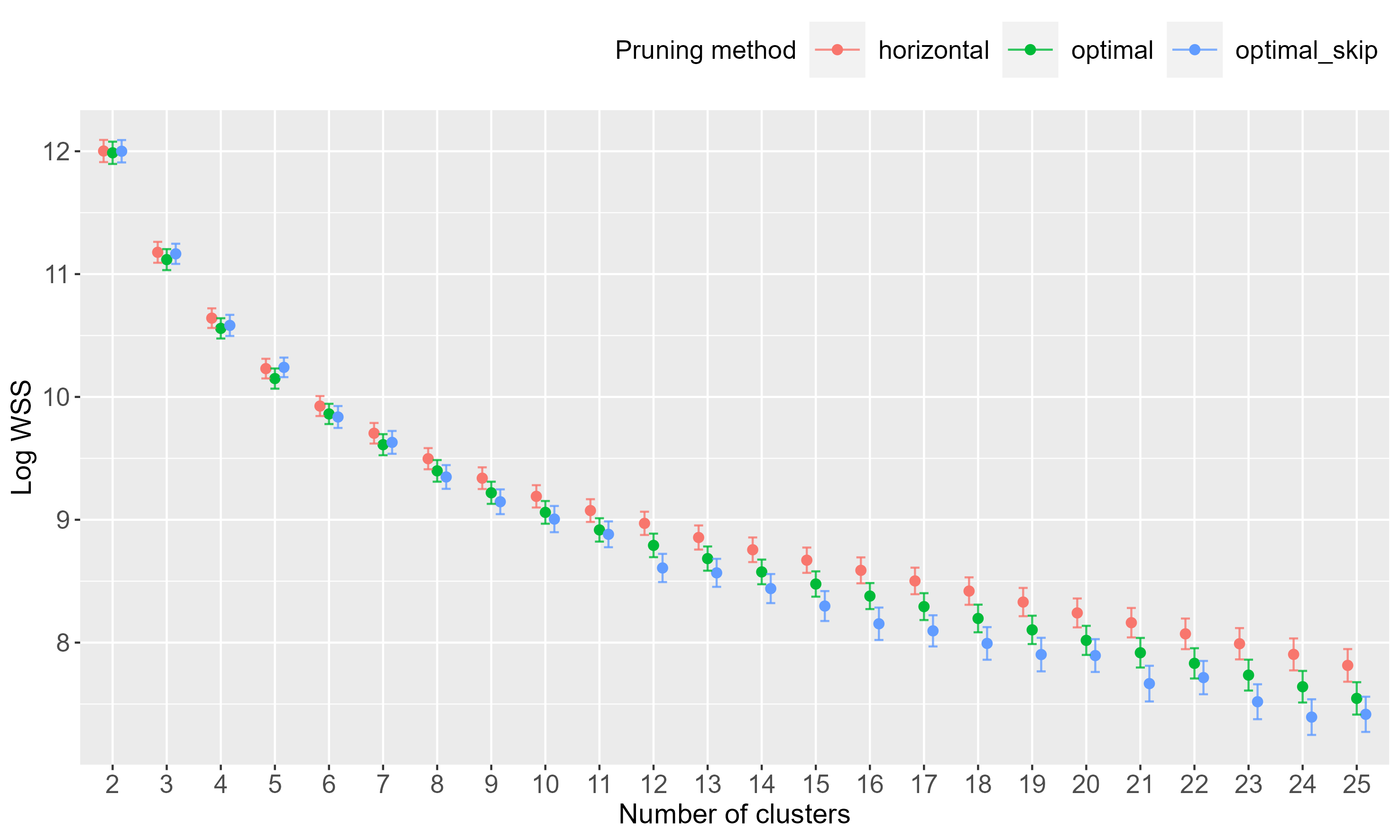}
    \caption{\em Log WSS mean and SE comparison of the two pruning methods on 200 random datasets with well-separated clusters. Optimal pruning increases the number of leaves to the closest available number if the subtree with a certain number of leaves is not in the sequence of optimal pruned subtrees, while Optimal\_skip skips the subtree.} 
    \label{vs2}
    \end{centering}
\end{figure}

We ran the two pruning methods, horizontal and optimal weakest-link pruning, on the 200 datasets. Figure \ref{vs2} shows the log of  WSS when the dendrogram was pruned down to between 2 and 25 terminal nodes. The colored circles are the mean of log WSS, and the error bar represents one standard error. For optimal pruning, if the subtree with a certain number of leaves was not in the sequence of optimal pruned subtrees, we increased the number of leaves to the closest available number. For the optimal\_skip pruning, if the subtree with a certain number of leaves was not in the sequence of optimal pruned subtrees, we skipped this subtree and thus the loss was set to NA. 

\begin{figure}
    \begin{centering}
    \includegraphics[scale=0.6]{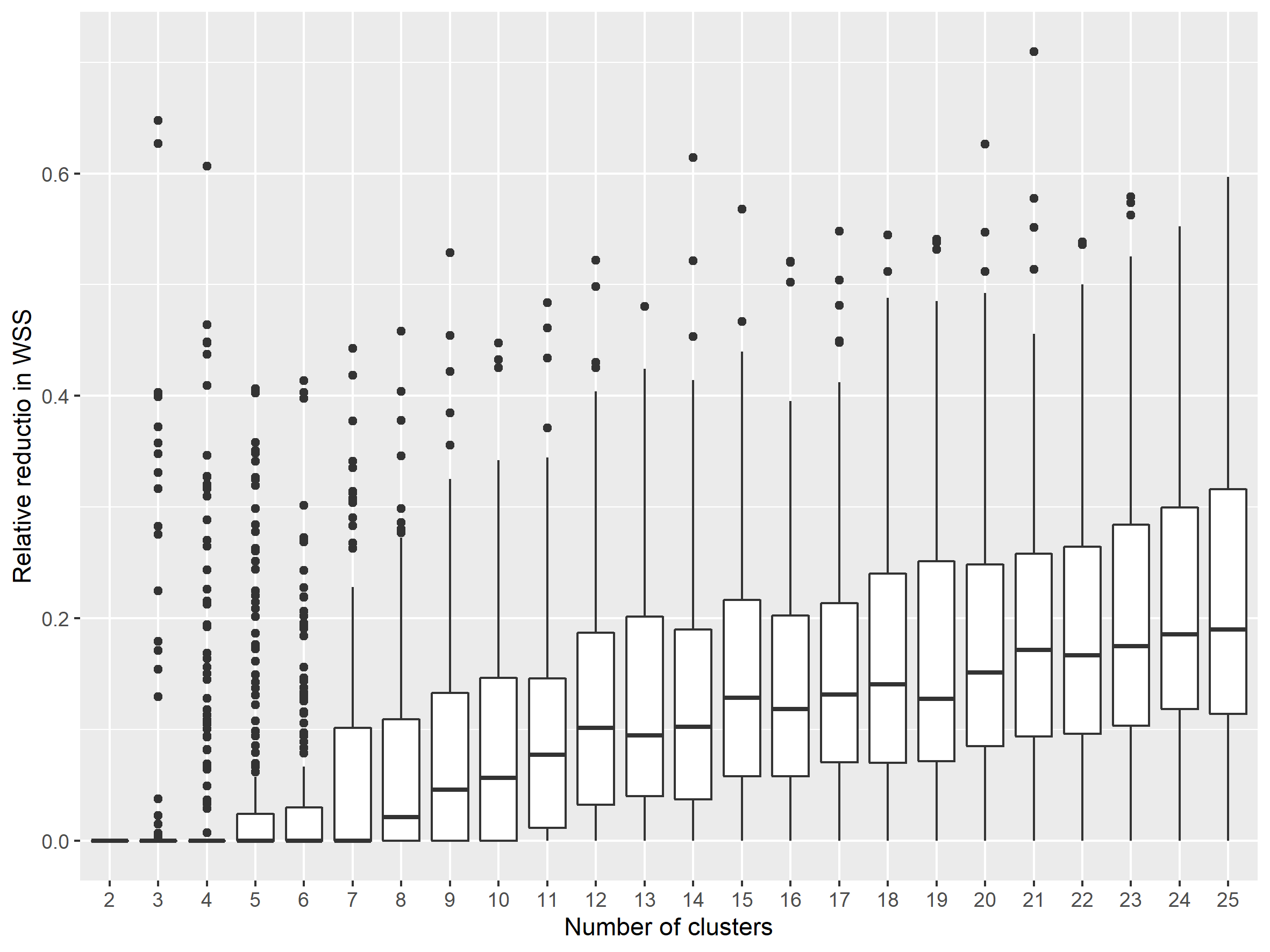}
    \caption{\em Relative reduction in WSS, for optimal pruning versus horizontal cut on 200 random datasets with well-separated clusters.} 
    \label{diff2}
    \end{centering}
\end{figure}

Figure \ref{diff2} shows the relative reduction in WSS for weakest-link pruning relative to horizontal cut. In this plot, the relative differences were only calculated when the subtree with the input number of clusters was a member of the sequence of minimizing subtrees. The results were similar to the previous results for the no-cluster setting, but the improvement is less dramatic. The median of relative reduction in loss (after removing NAs due to skipping numbers of clusters) was 0.085.

\subsection{Some real data examples}
Here we applied the pruning methods to six popular datasets from the book ``Introduction to Statistical Learning'', Second Edition (ISLR2) \citep{ISL}: Boston, Khan(train), Khan (test), NCI60, Portfolio, and USArrests. Details can be found in table \ref{table:real data}. 

\begin{table}[!ht]
\centering
\begin{tabular}{lcclll}
\multicolumn{1}{l|}{Dataset}      & \multicolumn{1}{l|}{Number of observations} & \multicolumn{1}{l}{Number of features} &  &  &  \\ \cline{1-3}
\multicolumn{1}{l|}{Boston}       & \multicolumn{1}{c|}{506}            & 13                            &  &  &  \\
\multicolumn{1}{l|}{Khan (train)} & \multicolumn{1}{c|}{63}             & 2308                          &  &  &  \\
\multicolumn{1}{l|}{Khan (test)}  & \multicolumn{1}{c|}{20}             & 2308                          &  &  &  \\
\multicolumn{1}{l|}{NCI60}        & \multicolumn{1}{c|}{64}             & 6831                          &  &  &  \\
\multicolumn{1}{l|}{Portfolio}    & \multicolumn{1}{c|}{100}            & 2                             &  &  &  \\
\multicolumn{1}{l|}{USArrests}    & \multicolumn{1}{c|}{50}             & 4                             &  &  &  \\
                                  & \multicolumn{1}{l}{}                & \multicolumn{1}{l}{}          &  &  &  \\
                                  & \multicolumn{1}{l}{}                & \multicolumn{1}{l}{}          &  &  &  \\
                                  & \multicolumn{1}{l}{}                & \multicolumn{1}{l}{}          &  &  & 
\end{tabular}
 \caption{\em Six real datasets from the ISLR2 text.}
 \label{table:real data}
\end{table}

\begin{figure}[!ht]
    \begin{centering}
    \includegraphics[scale=0.6]{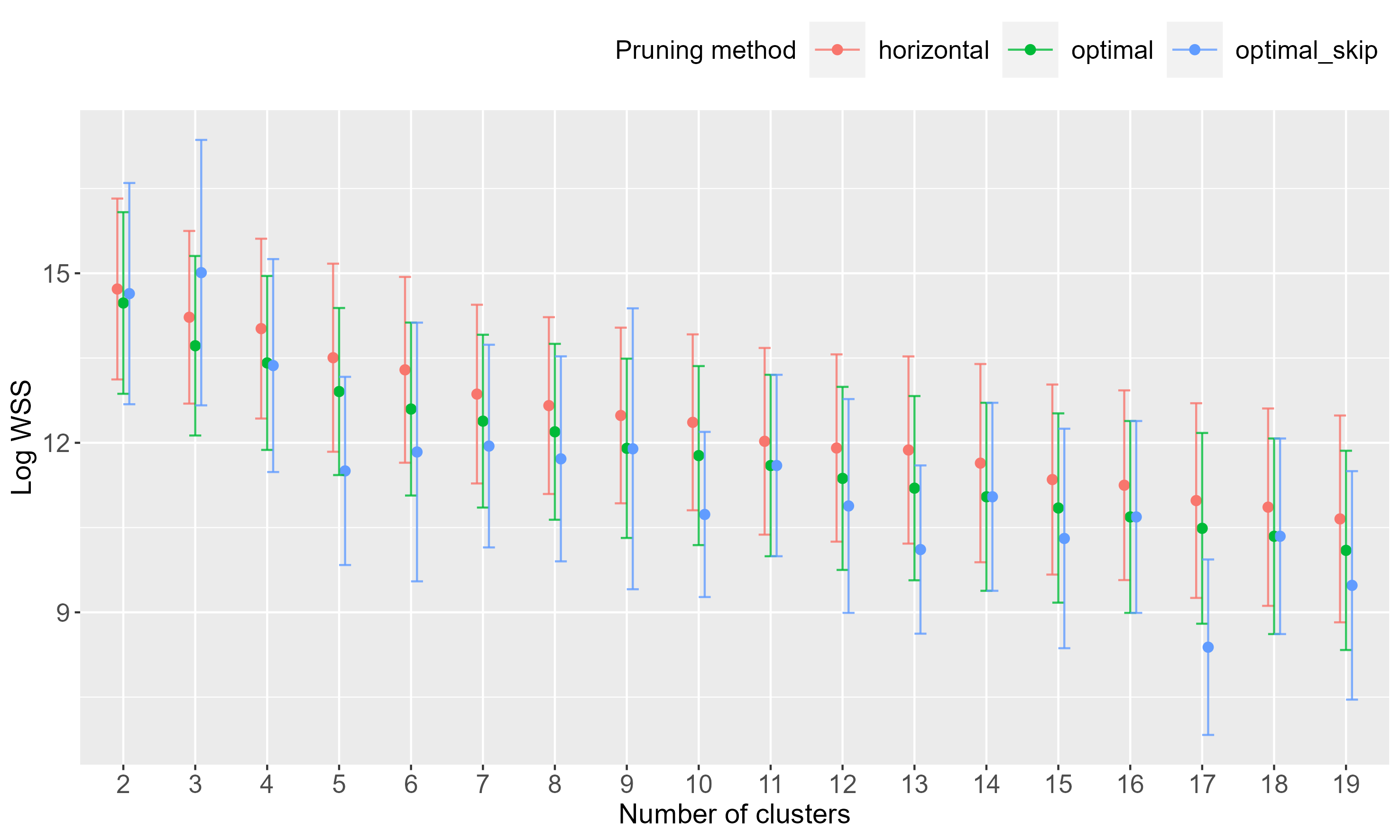}
    \caption{\em Log WSS mean and SE comparison of two pruning methods on six real datasets. Optimal pruning increases the number of leaves to the closest available number if the subtree with a certain number of leaves is not in the sequence of optimal pruned subtrees, while Optimal\_skip skips the subtree.} 
    \label{vs3}
    \end{centering}
\end{figure}

\begin{figure}[!ht]
    \begin{centering}
    \includegraphics[scale=0.6]{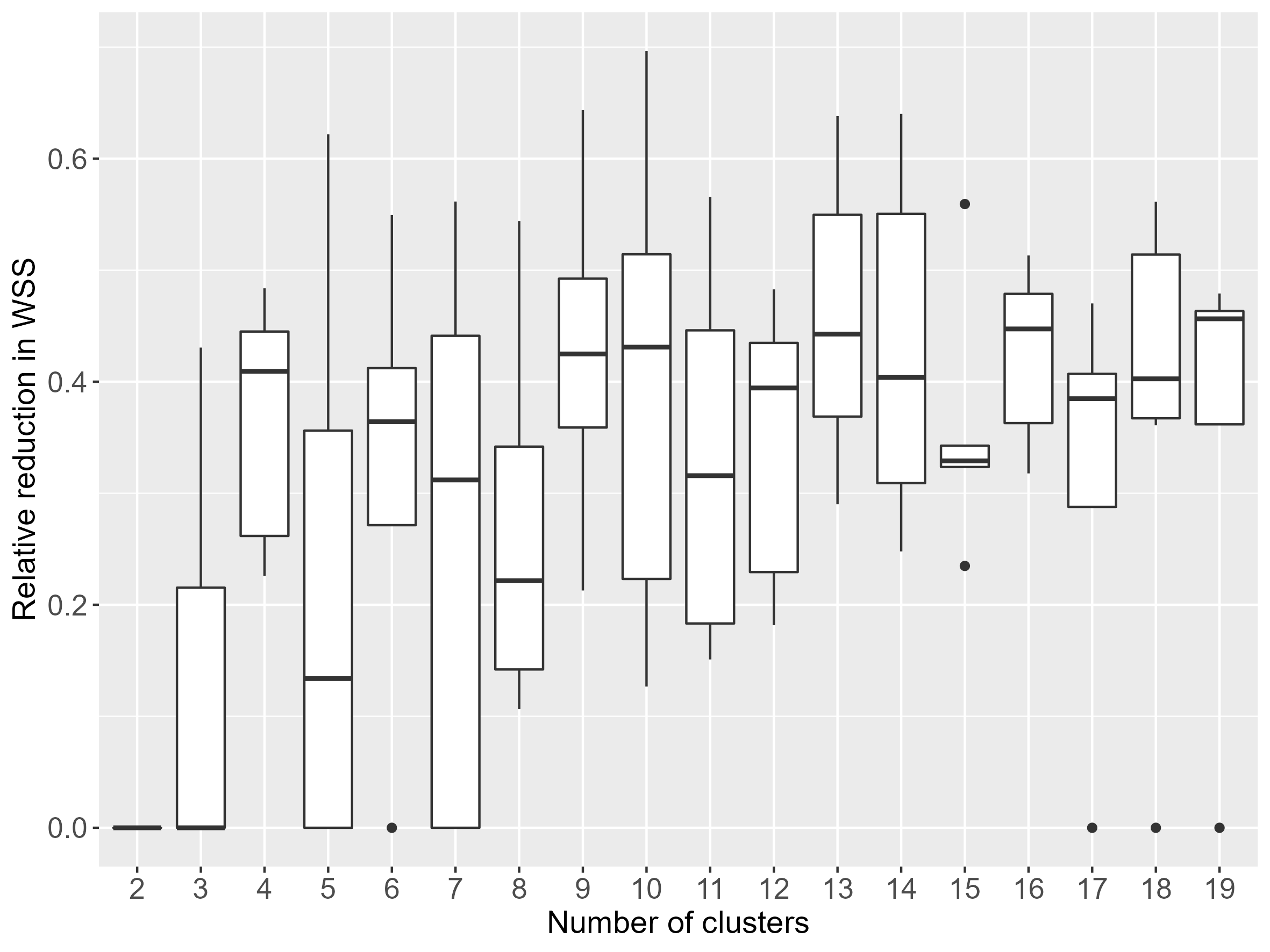}
    \caption{\em Relative reduction in WSS, for optimal pruning versus horizontal cut on 6 real datasets} 
    \label{diff3}
    \end{centering}
\end{figure}

We ran the two pruning methods on the six real datasets for comparison. Combining results for all six datasets together, Figure \ref{vs3} shows the log of WSS when each dendrogram was pruned to between 2 and 19 terminal nodes. The colored circles represent the mean of (log) WSS and the error bars represent one standard error. If the subtree with a certain number of leaves was not in the sequence of optimal pruned subtrees, we increased the number of leaves to the closest available number. Figure \ref{diff3} shows the relative reduction in loss when we replaced the horizontal cut with weakest link cut. In this plot, the relative differences were only calculated when the subtree with the input number of clusters was a member of the sequence of minimizing subtrees. The reduction in loss for optimal pruning versus horizontal cut was quite substantial.

\subsection{NCI60 classification Problem}
\label{sec:class}
The NCI60 dataset comes from cancer cell line microarrays, consisting  of 6,830 gene expression measurements on 64 cancer cell lines. There are 14 different cancer types in this dataset: "BREAST", "CNS", "COLON", "K562A-repro", "K562B-repro", "LEUKEMIA", "MCF7A-repro", "MCF7D-repro", "MELANOMA", "NSCLC", "OVARIAN", "PROSTATE", "RENAL", "UNKNOWN". First ignoring the cancer labels, we built a dendrogram using average linkage, and performed optimal and horizontal pruning methods, pruning down to 14 terminal nodes in each case. 

\begin{figure}[!ht]
\centering
  \includegraphics[scale = 0.5]{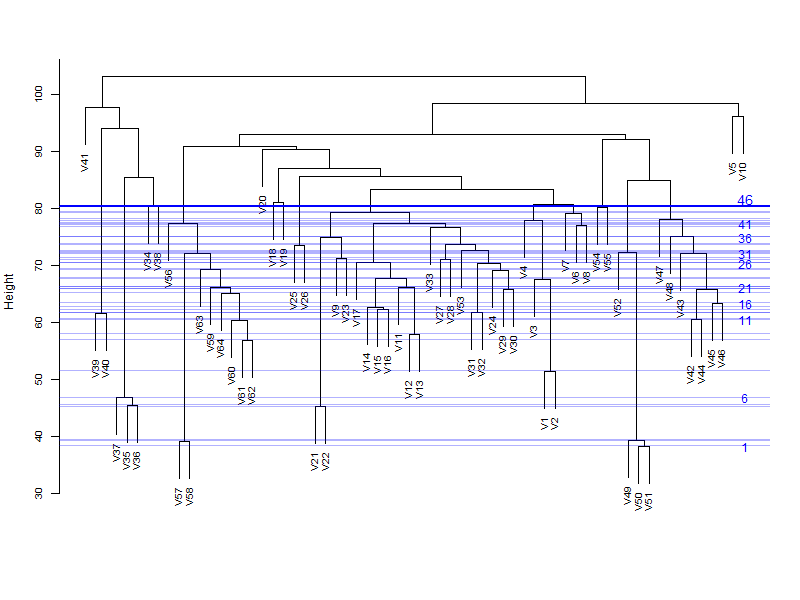}
  \caption{\em The horizontal pruning steps resulting in 14 leaves for NCI-60}
  \label{fig:class_ex_hc_h}
\end{figure}

\begin{figure}[!ht]
\centering
  \includegraphics[scale = 0.5]{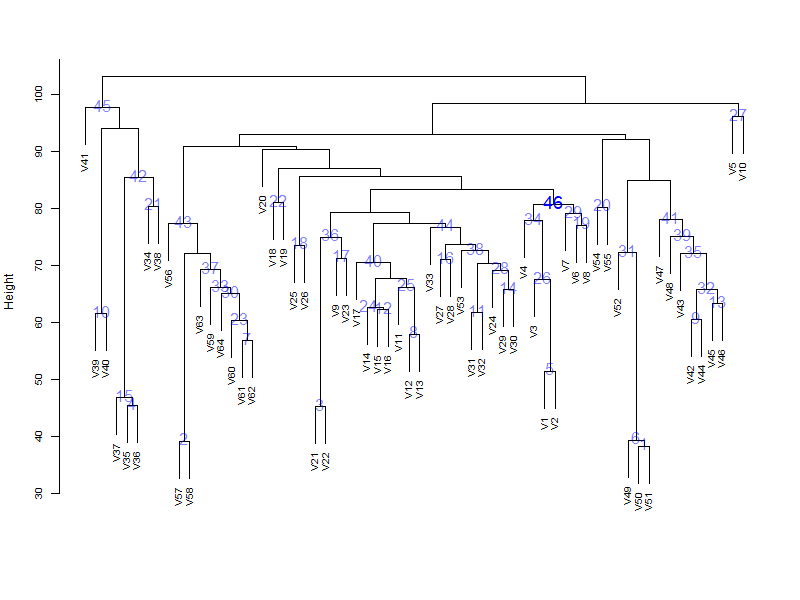}
  \caption{\em The optimal pruning steps resulting in 14 leaves for NCI-60}
  \label{fig:class_ex_hc_o}
\end{figure}

Figure \ref{fig:class_ex_hc_h} shows the sequence of horizontal cuts. We label every 5 cuts with transparent thin blue lines and the final cut with the solid thicker blue line. Figure \ref{fig:class_ex_hc_o} shows the sequence of optimal cuts. We label cuts with transparent blue numbers and the last cut with the bigger solid blue ``46''.

\begin{figure}[!ht]
\centering
  \includegraphics[scale = 0.45]{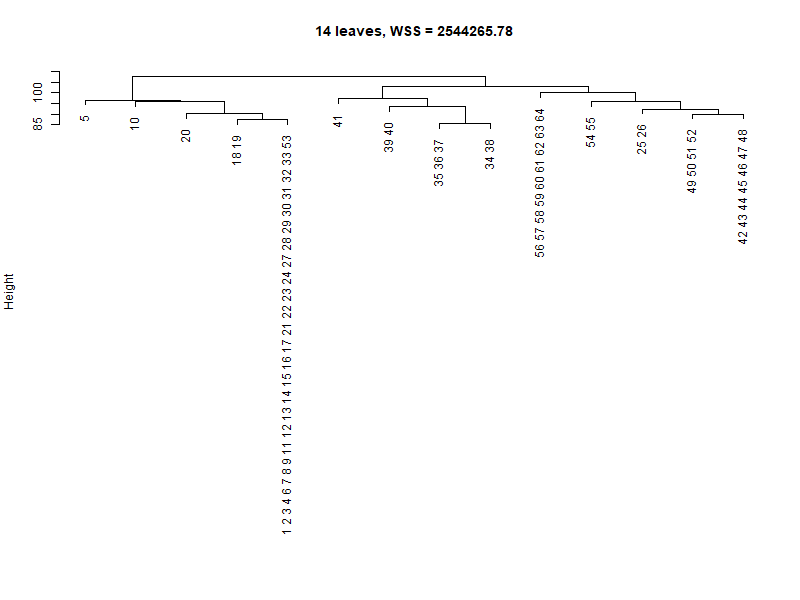}
  \caption{\em NCI-60: The horizontal pruned tree with 14 leaves }
\label{fig:NCI_h_14}
\end{figure}

\begin{figure}[!ht]
  \centering
  \includegraphics[scale = 0.45]{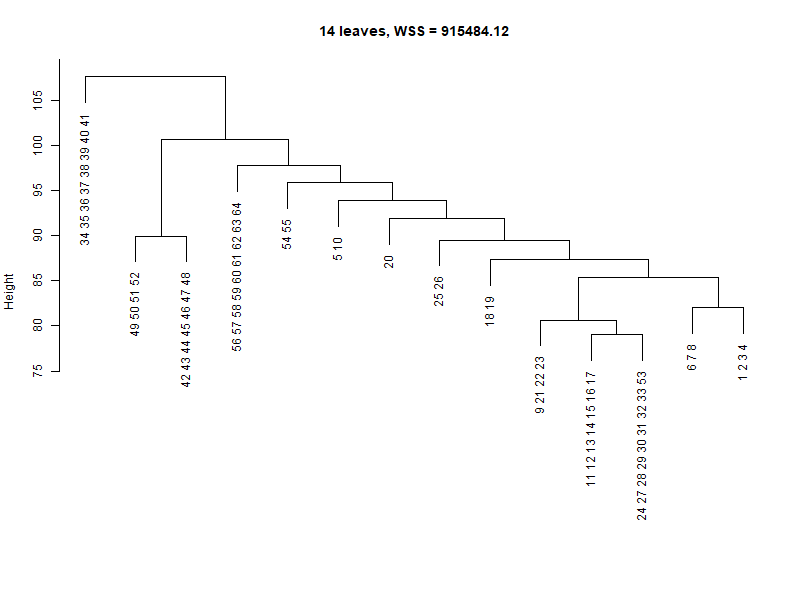}
  \caption{\em NCI-60: The optimal pruned tree with 14 leaves.}
\label{fig:NCI_o_14}
\end{figure}

\begin{figure}[!ht]
\centering
\begin{subfigure}{.5\textwidth}
  \centering
  \includegraphics[scale = 0.4]{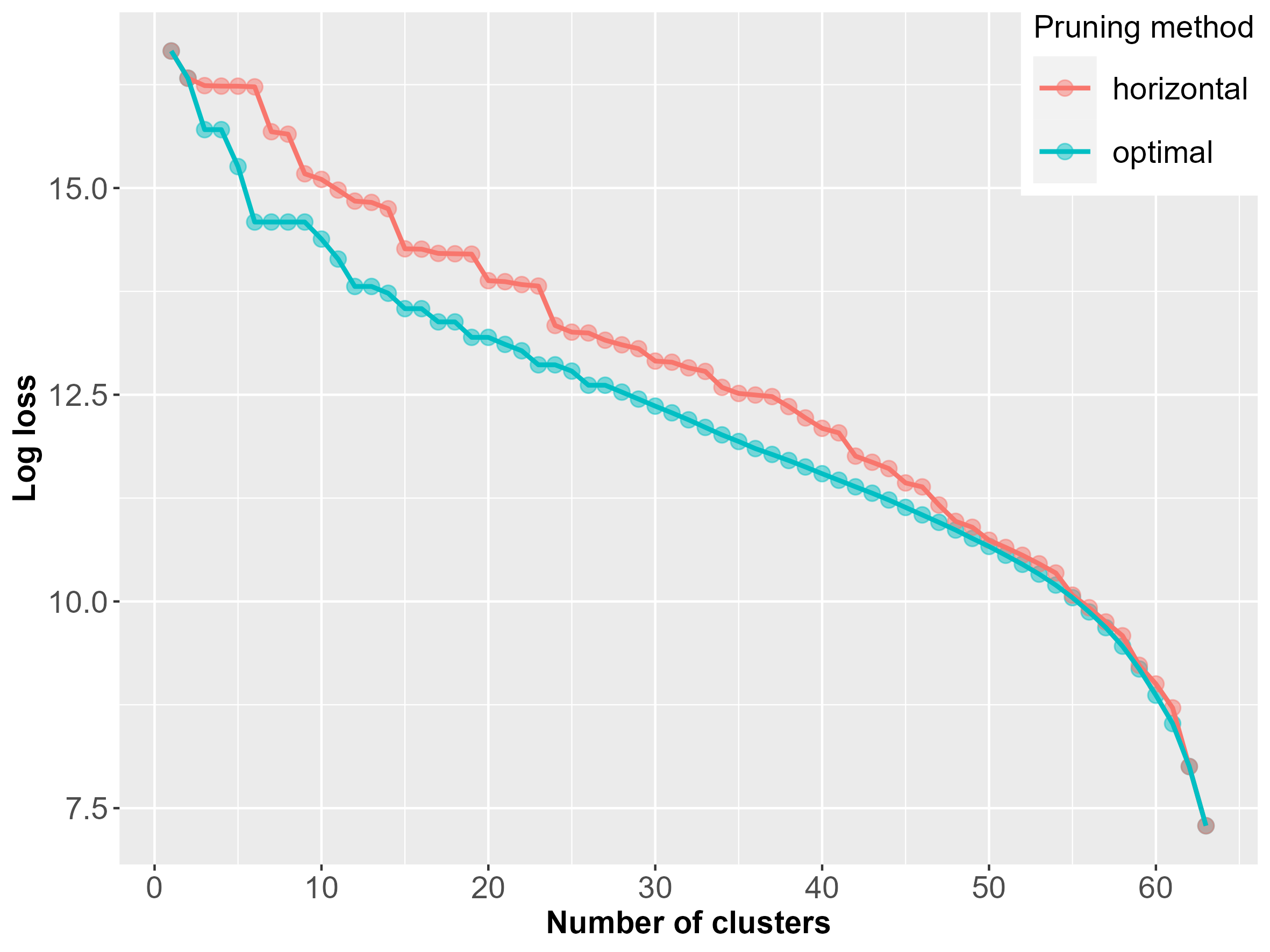}
  \caption{\em  WSS comparison}
\end{subfigure}%
\begin{subfigure}{.5\textwidth}
  \centering
  \includegraphics[scale = 0.4]{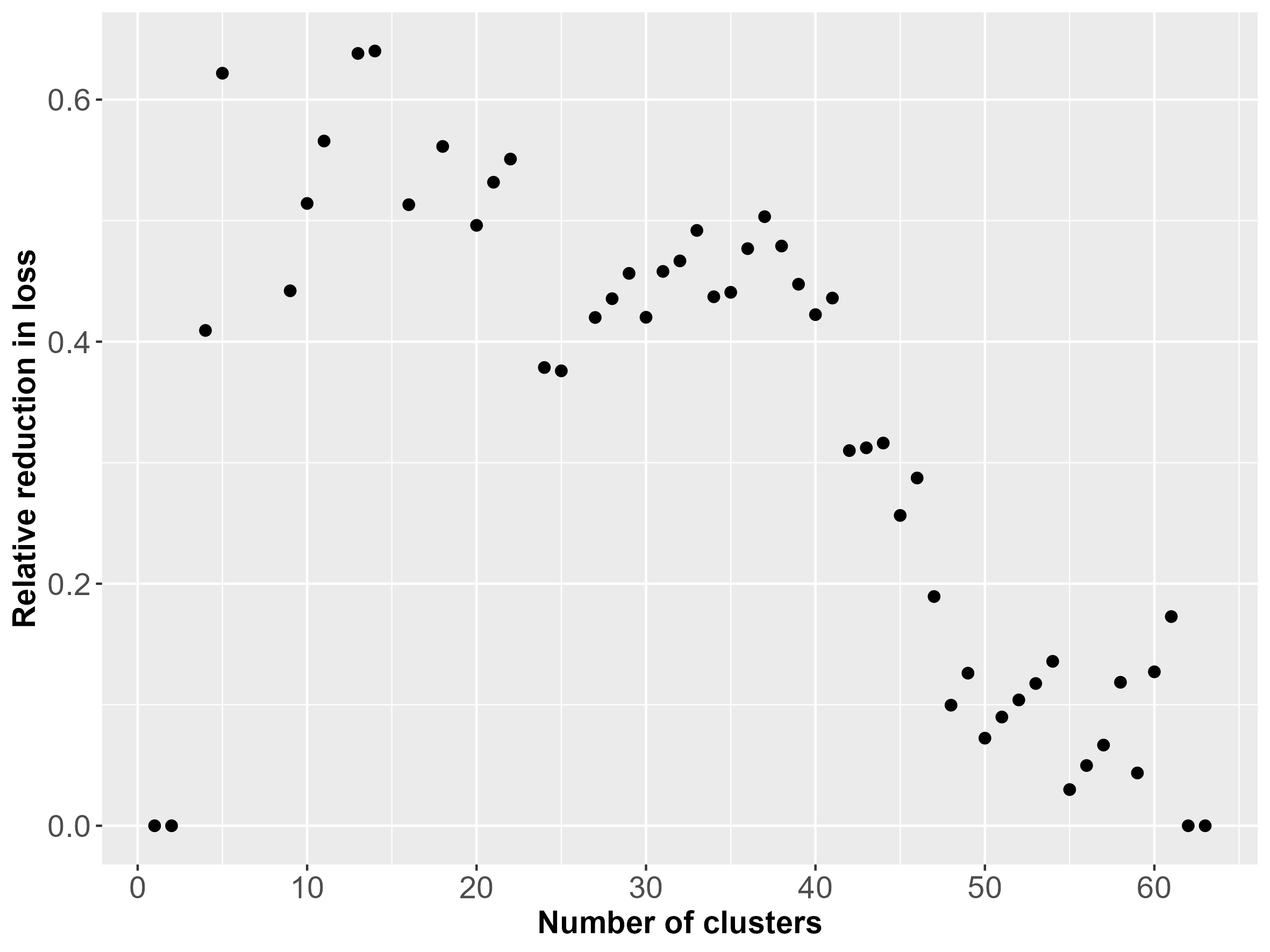}
  \caption{\em Relative reduction in WSS}
\end{subfigure}
\caption{\em Comparison of two pruning methods NCI-60}
\label{fig:NCI}
\end{figure}

The final dendrogram produced by horizontal pruning is shown in figure \ref{fig:NCI_h_14} and the optimal pruned one is shown in figure \ref{fig:NCI_o_14}. The WSS for the horizontal pruned tree was $2,544,265.78$ while the final WSS for the optimal pruned tree was $915,484.12$, around only one-third of the value from the horizontal counterpart. The median relative reduction in WSS (after removing NAs due to skipping numbers of clusters) was 0.36. As we can see from Figure \ref{fig:NCI}, optimal cut dominates horizontal cut for all numbers of clusters and the discrepancy is more significant when the number of clusters is between 5 and 40. 

\begin{figure}[!ht]
\centering
\begin{subfigure}{.5\textwidth}
  \centering
  \includegraphics[scale = 0.15]{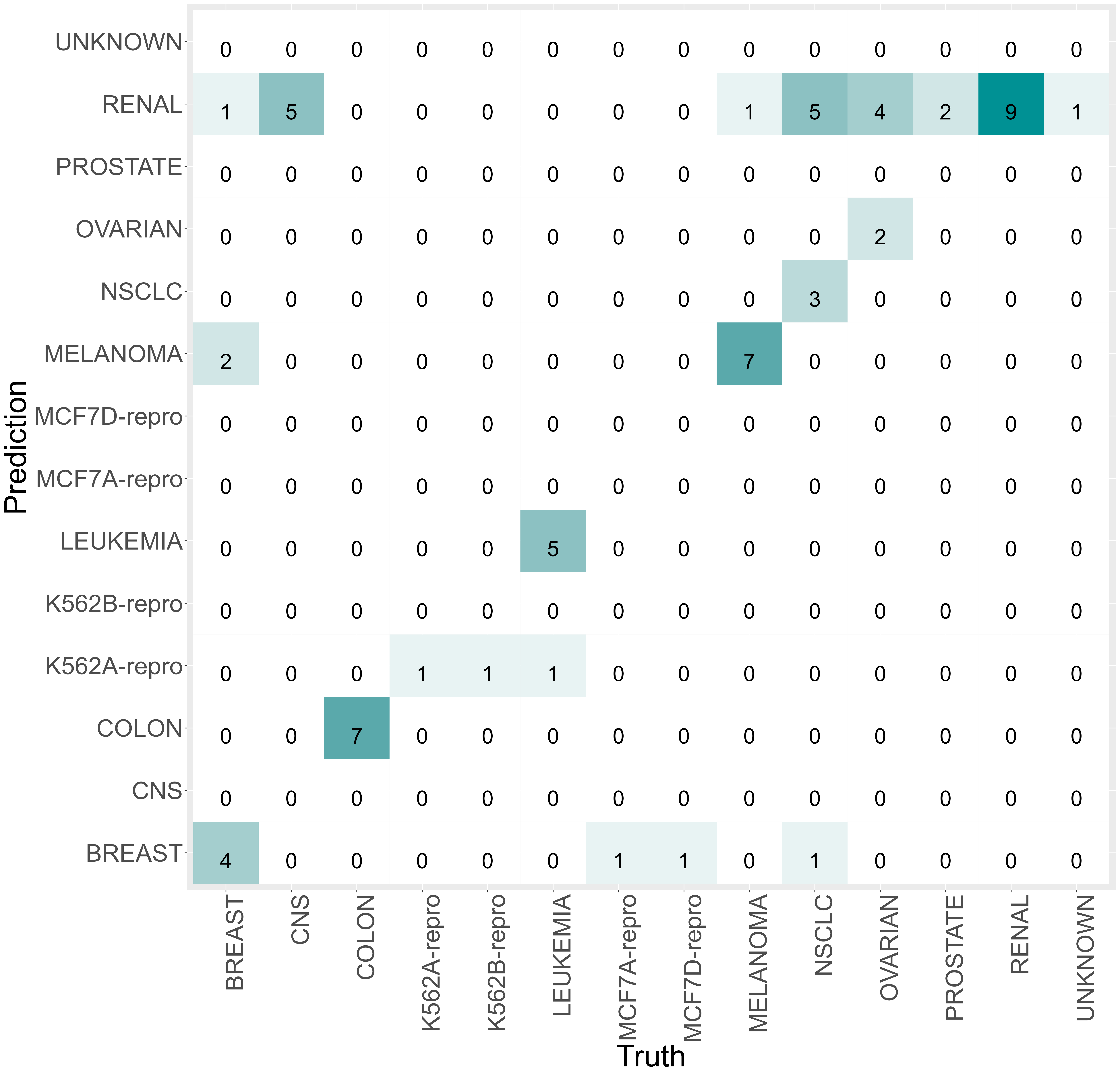}
  \caption{\em Using horizontal pruning}
\end{subfigure}%
\begin{subfigure}{.5\textwidth}
  \centering
  \includegraphics[scale = 0.15]{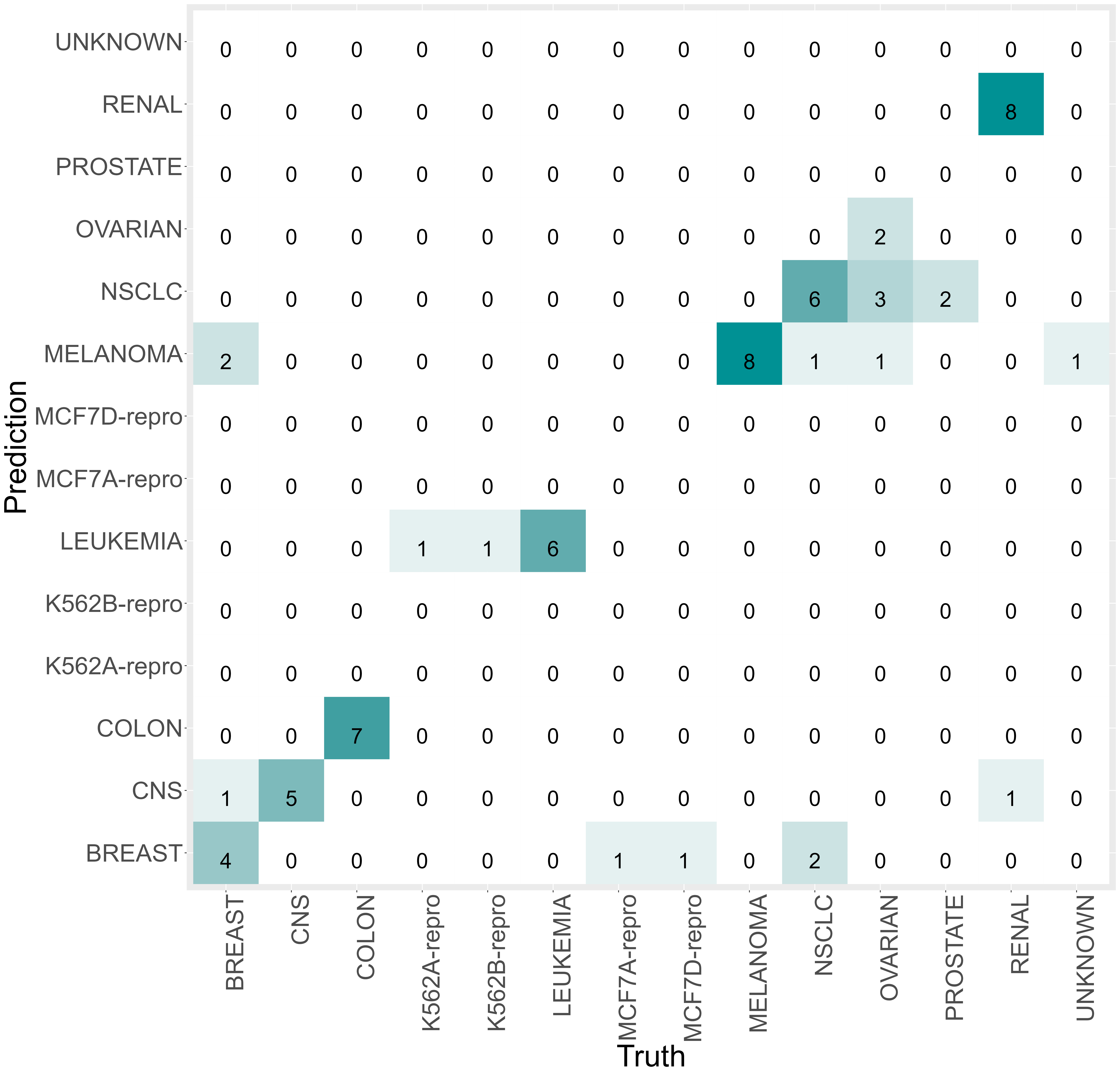}
  \caption{\em Using optimal pruning}
\end{subfigure}
\caption{\em Confusion matrices for classifying 14 cancer cell lines }
\label{fig:cm}
\end{figure}

Finally, for both pruning methods, we focused on the resulting trees with 14 terminal nodes (the number of cancer classes). We used a majority vote of the cell lines in each cluster to determine a predicted cancer label for that cluster. We then compared these predictions to the true sample labels. The confusion matrices are shown in Figure \ref{fig:cm}. The classification error rate was 0.41 for the horizontal pruned tree, and 0.28 for the optimal pruned tree. The results suggest that our optimal pruning has yielded a more meaningful grouping of the observations. 

\section{Choosing the best-sized subtree}
\label{sec:gap}

With the sequence of optimal subtrees in hand, we can estimate the best-sized subtree among the sequence. This is equivalent to choosing the optimal value of the cost-complexity parameter $\alpha$. In CART, this is done via cross-validation, but this approach is not available in the unsupervised setting. So instead we apply a method for choosing the optimal number of clusters. There are many such methods, with two popular ones being the silhouette test (Rousseeuw) \citep{silhouett} and the Gap test (Tibshirani et al.) \citep{gap}. For illustration, we use the Gap test here. 

The Gap test compares the changes in within-cluster dispersion to what would be expected under a suitable reference null distribution using the output of any clustering algorithm, such as hierarchical clustering. More specifically, it compares the curve of log  within-cluster dissimilarity against the number of clusters to the curve achieved from data uniformly distributed throughout a rectangular space. The optimal number of groups can be chosen based on many criteria, here we chose the place where the gap between the two curves is most significant.

\begin{figure}[!ht]
\centering
  \includegraphics[scale = 0.6]{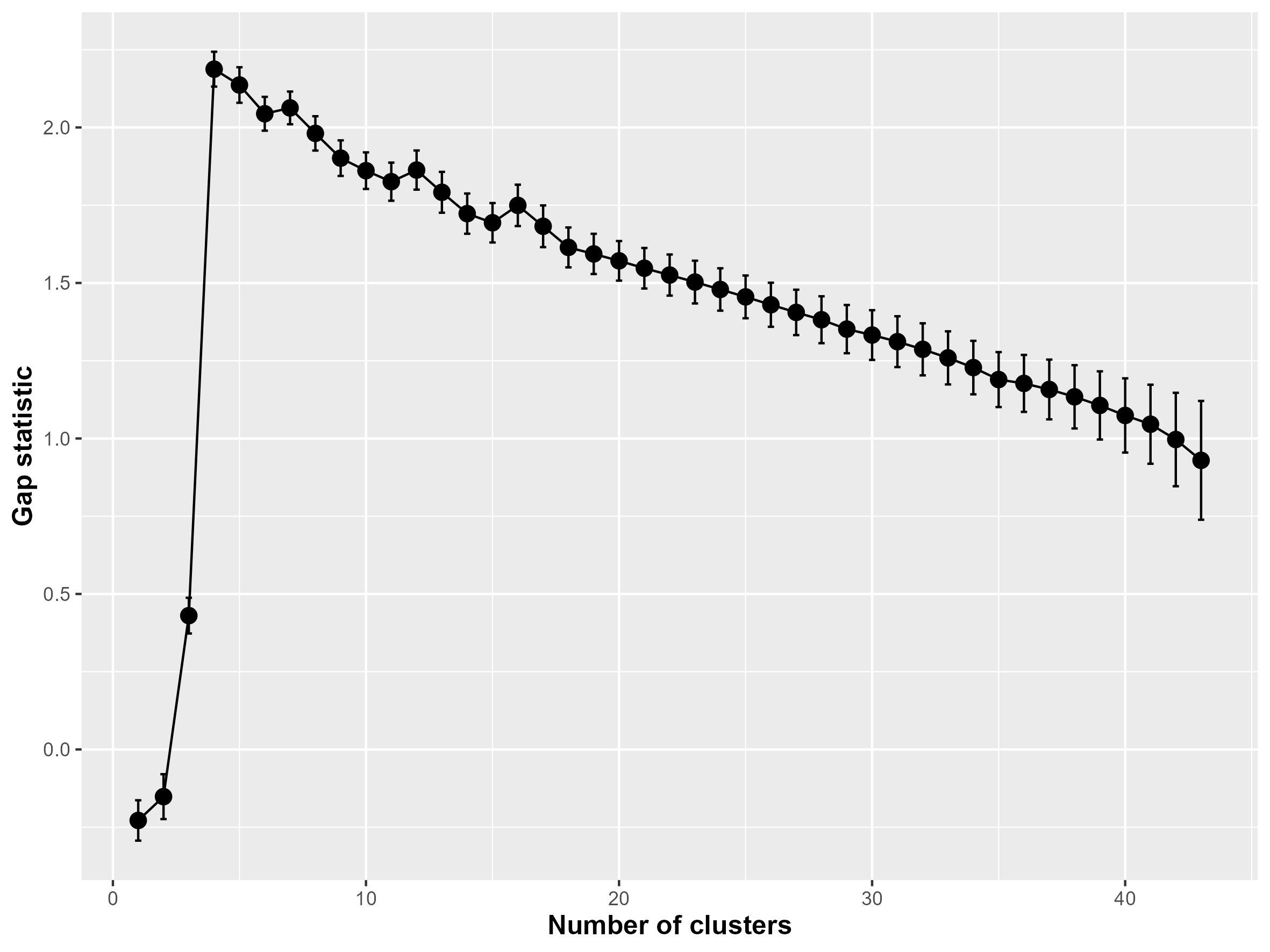}
  \caption{\em A Gap test example of 44 data points and 25 dimensions. The estimated number of clusters is 4, the true number.}
  \label{fig:gap}
\end{figure}

We randomly generated 20 datasets $\mathit{X_i}$ with the number of observations $n_i$  between 20 and 30, the number of features $p_i$ between 1 and 30, and the number of well-separated clusters equal to 4. We sampled both $n_i$ and $p_i$ uniformly. Then we used the {\tt monte} function in the R package {\tt fungible} \citep{fungible} to generate the $x_{ij}$. The sizes of all clusters were equal, except for the last cluster which adjusted for indivisibility. Then we ran the Gap test with the optimal pruning method on the 20 datasets. If the subtree with a certain number of leaves was not in the sequence of optimal pruned subtrees, we increased the number of leaves to the closest available number. The Gap test chose the correct number of clusters (4) 18 out of 20 times. Figure \ref{fig:gap} shows an example that the optimal number of clusters was 4. Of course, the Gap test (and other tests) will not perform as well when the clusters are not well separated.

\section{Discussion}
\label{sec:discussion}
In this paper, we have proposed the pruning of a dendrogram by a weakest-link approach. This gives a sequence of the tightest clusters, for each fixed number of groups. This sequence can be used in an exploratory way or can be combined with a procedure such as the Gap test to estimate the best set of clusters.

A public domain R language package {\tt pruneClust} 
will soon be available on the CRAN repository.

\medskip

{\bf Acknowledgments}. We thank Jacob Bien for helpful comments, and Tal Galili
for help with the dendextend package R package. RT is funded by the National Institutes of Health (5R01 EB001988-16) and the National Science Foundation (19 DMS1208164)

\bibliographystyle{plain} % We choose the "plain" reference style
\bibliography{reference}
\end{document}